\documentclass{cimento}

\usepackage{amsmath,amssymb,epsfig}

\newcommand{\be}{\begin{equation}}
\newcommand{\ee}{\end{equation}}
\newcommand{\bea}{\begin{eqnarray}}
\newcommand{\eea}{\end{eqnarray}}
\newcommand{\bean}{\begin{eqnarray*}}
\newcommand{\eean}{\end{eqnarray*}}
\newcommand{\bc}{\begin{center}}
\newcommand{\ec}{\end{center}}

\newcommand{\CM}{{\cal M}}
\newcommand{\cX}{{\cal X}}
\newcommand{\IP}{\mathbb{P}}

\newcommand{\f}{{\cal F}^{\flat}}
\newcommand{\firr}[1]{{}^{{\rm Irr}}\!{\cal F}^{\flat}_{#1}}

\title{Operators and vacua of $\mathcal{N}=1$ field theories}
\author{Davide Forcella\from{ins:x}\ETC,}
\instlist{\inst{ins:x} CNRS et Laboratoire de Physique Theorique de l'Ecole Normale Superieure \\
24, rue Lhomond, 75321 Paris Cedex 05, France\\
 forcella@lpt.ens.fr }
\PACSes{PACS: 11.25.Tq, 11.25.Uv, 11.25.Hf}

\begin{document}

\maketitle
\begin{flushright}
LPTENS-09/41\\
\end{flushright}
\begin{abstract}
We review the idea of Hilbert Series as a 
tool to study the moduli space and the BPS operators 
of four dimensional $\mathcal{N}=1$ supersymmetric field theories.
We concentrate on the particular case of 
$\mathcal{N}=1$ superconformal field theories 
living on N D3 branes at toric Calabi-Yau singularities.
The main claim is: it is possible to write down explicit 
partition functions counting all the local BPS operators for generic N number of branes,
 and obtain important informations about the BPS operators, the moduli space and the
dual geometry. 
\end{abstract}



\section{Introduction}
 
In this paper I would like to give an overview of the study of supersymmetric vacua 
and BPS operators of $\mathcal{N}=1$ supersymmetric field theories using the idea 
of Hilbert Series (HS). 
I will consider as explicit example the super-conformal field theories (SCFT) 
living on a stack of N D3 branes at the tip of the Calabi-Yau (CY) six dimensional 
conical singularity $\cX=C(H)$. 

This kind of ``new'' approach to $\mathcal{N}=1$ supersymmetric field theories was born from a
set of very simple observations: every $\mathcal{N}=1$ supersymmetric field theory has a 
special subset of operators: the local BPS operators, we would like to have a generating function 
(HS) counting them according to their global charges; 
the local BPS operators are the holomorphic functions 
over the moduli space: once we know the BPS spectrum of the theory 
we also know its moduli space.

To write down explicit HS counting all the local BPS operators of a
non-abelian field theory is a rather complicated task and we will refer the interested 
reader to the original literature \cite{Kinney:2005ej,Martelli:2006yb,Benvenuti:2006qr,Butti:2006au,
Feng:2007ur,Forcella:2007wk,Butti:2007jv,Forcella:2007ps,Forcella:2008bb,Forcella:2008ng}. 
In this review paper we would just like to point out that the HS contains 
a lot of informations related to the field theory, and it is worthwhile to study them. 

In particular the HS contains the structure of generators and relations 
of the chiral ring and it knows about the dimension, the symmetries 
and the algebraic geometric
structure of the complete moduli space.

In the special case of SCFT living on D-branes at singularities we can calculate the HS at 
weak coupling, using field theory, or at strong coupling using string theory
and compare the two results. Moreover the HS knows about the central charge and the exact R-charges 
of the elementary fields, it contains part of the KK spectrum in supergravity and the values of
the volume of the supergravity compactification manifold H and of some of its supersymmetric three cycles. 

With this short introduction we hope to have motivated the reader to know more about 
HS. In the following we will at first introduce our favorite 
example of supersymmetric theories: the SCFT living on N D3 branes 
at the tip of the CY singularity $\cX=C(H)$; we will then divide the study 
of their chiral ring in $N=1$ and $N>1$: we will introduce the idea of Master Space (MS) 
and we will illustrate the generic structure of the HS for N=1; we will comment on some 
of the properties of the MS and will summarize some of the informations we could obtain 
from the HS. We will then study the $N>1$ case: we will introduce the idea of Plethystic Exponential 
(PE) and we will write down the generic form of the HS counting all the local BPS operators for 
generic N for every SCFT on D3 branes at toric CY singularity; we will comment on some of its 
properties and we will explain some of the informations we can obtain from the HS regarding the complete
moduli space of the theory for generic N. We will finally analyze, with an explicit example, some
of the generalities explained all along the paper, and we will conclude with some comments about
other possible applications.
\newline
\newline
This paper is based on the author's PhD thesis \cite{Forcella:2009bv}, 
presented at the INFN conference, 21/09/2009, Rome, for the "Premio Fubini 2009". 
See also \cite{string2009}.

\section{D3 branes at singularities}

In this paper we will concentrate on the SCFT living on N D3 branes at the tip of 
 CY singularity $\cX=C(H)$.
The main reasons for this choice are: they are simple enough to obtain explicit 
and general results, and they are very interesting for at least two reasons: 
they are the basis of very interesting phenomenological models that embed generalizations of
the MSSM and SUSY breaking mechanisms in string theory; and they all have a strong coupling description in the AdS/CFT correspondence: 
type IIB string theory on $AdS_5 \times H$.

The low energy dynamics of D3 branes at $\cX$ is a quiver gauge theory 
\cite{Douglas:1996sw}: it is an $\mathcal{N}=1$
supersymmetric field theory with a manifold of IR superconformal fixed points, it has
 gauge group $\prod_i^g SU(N_i)$, chiral bifoundamental 
matter fields, and a particular form 
for the superpotential. 
For simplicity we will always assume that the singularity $\cX$ is toric, meaning that it 
admits at least a $U(1)^3$ isometry. In this case there is a rather complete dictionary 
between the field theory and the dual geometry.

The natural presence of chiral fields is what makes these theories phenomenologically 
interesting. Indeed generalizing a bit this setup 
one can reproduce very interesting MSSM-like scenario, SUSY-breaking, renormalization group flows and
confinement, moreover they are nice models to study non-perturbative instanton processes 
that could produce some perturbatively forbidden couplings like some mass terms or Yukawa interactions. 

On the other end the AdS/CFT correspondence claims that the field theory living on 
N D3 branes at the $\cX=C(H)$ singularity is dual, 
in a strong/weak coupling sense,  
to type IIB string theory propagating on the $AdS_5 \times H $ 
background with N unit of flux: $\int_H F_5 = N $.

We would like to study two main objects of these field theories: 
their moduli spaces and their BPS operators for generic N number 
of D branes. The moduli space $\mathcal{M}_N$ is the variety of expectation 
values of the scalar fields that do not break the supersymmetry;
the local BPS operators are local gauge invariant operators $\mathcal{O}$ 
killed by half of the supercharges $\bar{D} \mathcal{O} = 0 $.

The gauge group factors are SU(N) groups and they admit two kinds of invariant
tensors: $\delta^i_j$ and $\epsilon_{i_1,...,i_N}$. The gauge invariant operators constructed contracting the
elementary bifoundamental fields with the $\delta$ 
are trace-like and are called mesonic operators, while the operators obtained using the $\epsilon$ 
are determinant-like and are called baryonic operators. These are the two kinds of BPS operators 
we want to study. 

The global symmetries of these SCFTs come from the geometry of $\cX$. There exist two 
different types of symmetries: the mesonic ones coming from the isometries of $\cX$ 
and the baryonic ones under which only the determinant like operators are charged and 
coming from the topology of three cycles of H. 

The setup of D branes at singularity has a set of nice properties 
that will make easier our analysis. A first crucial observation 
is that at least a part of the moduli space of the gauge theory 
is explicitly realized in the geometry. 
Indeed some of the scalar degrees of freedom parameterizing 
the moduli space $\CM$ must also parameterize the space transverse to 
the branes. In the abelian case we just have one D3 brane and it is just 
a free point in the transverse geometry, hence the moduli space of the theory 
must contain at least the six dimensional singularity $\cX$. 
In the generic non-abelian case we have N D3 branes that, 
being mutually supersymmetric, will behave as a set on N non-interacting particles on $\cX$. 
Because the ground state of the D3 branes is bosonic we have a system of N bosons 
and the non-abelian moduli space must contain the N-times symmetric product of the 
transverse singularity: Sym$^N(\cX)= \cX^N / \Sigma_N$, where 
$\Sigma_N$ is the group of permutations of N objects acting on the N factors $\cX$.

These nice properties allow for geometrical intuition and 
subdivide the problem in N=1 and $N>1$.

\section{Hilbert Series and Master Space: the N=1 case}

Our path to understand the complete spectrum of local 
BPS operators and the full moduli space $\CM_N$ will be 
somehow an unusual path: we want to compute HS!

The set of holomorphic functions over the moduli space $\CM$ is exactly the set of all local
BPS operators of the gauge theory. We want to construct a function that encode the 
informations of the chiral ring of the theory. We proceed in the following way. 
The gauge theory has generically $g+2$ global abelian symmetries: 
$U(1)^2 \times U(1)_R$ coming from the isometries of $\cX$, the mesonic symmetries, and 
$U(1)^{g-1}$ baryonic symmetries: anomalous and non-anomalous, 
coming from the $U(1)$ factors inside the $U(N_i)$ of the UV description of the quiver 
gauge theory that decouple in the IR superconformal fixed points; the ones associated 
to the three cycles in H are the non-anomalous ones.
 
We want to count the BPS operators according
to their charges under these various $U(1)$s. 
Let us introduce a set of $g+2$ chemical potentials $t_i$ with $i=1,...,g+2$. 
We would like to have a function that once expanded for small values of the 
chemical potentials 
\begin{equation}\label{HSintro}
H(t;\CM)=\sum_{i_1,...,i_{g+2}}c_{i_1,...,i_{g+2}}t_1^{i_1}...t_{g+2}^{i_{g+2}}
\end{equation}
gives the numbers $c_{i_1,...,i_{g+2}}$ of BPS operators with charges 
$i_j$ under the corresponding global abelian symmetries. 
In the mathematical literature this function is called the Hilbert Series of $\CM$, and
 it is actually counting the holomorphic function over the algebraic variety $\CM$ 
according to their charges under the action of its symmetry group.

In general $\CM$ has a nice symplectic quotient description: it is the quotient of the
algebraic variety $\f$ defined by the derivatives of the superpotential with respect to the elementary
fields (F-flatness conditions) by the complexified gauge group $G$:
\begin{equation}\label{M}
 \CM \simeq \f / G \ .
\end{equation}
This description of the moduli space clearly individuate a parent space $\f$, that we will
call the Master Space (MS) of the theory \cite{Forcella:2008bb}, and it will be very useful to study the full 
moduli space $\CM_N$ for generic N. The MS is a gauge dependent space that can be associated to
every $\mathcal{N}=1$ field theory, it is simpler than the full moduli space and we could start enjoying 
to classify its properties. In the case of D3 branes at $\cX$ it turns out that the knowledge of
just a subset of the MS for one brane is enough to obtain the HS counting BPS operators for generic N!

Let us start discussing the abelian N=1 case. The plan will be the following:
 we will fully characterize the moduli 
space in the abelian case and we will then explain the generic structure 
of the HS (\ref{HSintro}) for one brane. In the next section we will discuss
the non-abelian case.
 
With just one brane we are left with a set of $SU(1)$ groups, 
the gauge dynamics is trivial and $\CM_1=\f_1$. 
The MS $\f_1$ is the complete moduli space of the theory 
and contains as a sub set (the locus where all the baryonic operators 
have zero vacuum expectation value) the geometrical six dimensional 
transverse singularity $\cX$.

The master space is a $(\mathbb{C}^*)^{g-1}$ fibration over 
$\cX$ with the fiber structure induced by the baryonic symmetries. It is 
nice to observe that $\f_1$ has a very rich structure:
\begin{itemize}
\item{$\CM_1=\f_1$ is a $g+2$ dimensional toric variety, 
and can be completely characterized in term of algebraic geometry. 
In particular $\f_1$ is generically reducible in 
\begin{itemize}
\item{a $g+2$ dimensional component $\firr{1}$ that is a toric Calabi Yau cone,} 
\item{and a set of lower dimensional generically linear components $L_i$.}
\end{itemize}
}
\end{itemize} 

It is possible to show that $\firr{1}$ has the rigid structure already explained and moreover 
it has nice transformation properties under Seiberg dualities and a defined meaning in 
the dual string theory/strongly coupled description. A case by case
analysis shows instead that the lower dimensional spaces $L_i$ are less constrained, 
even if one can show that they could parameterize a subclass of RG flows of the theory. 
For these reasons we would like to concentrate on $\firr{1}$, that indeed 
turn out to be the variety containing the necessary informations to obtain the 
BPS spectrum for generic N.

We know that $\firr{1}$ is the fibration: $(\mathbb{C}^*)^{g-1} \rightarrow \mathcal{X}$, where
the fiber are generated by the baryonic symmetries of the theory, while the isometries 
of the base space $\mathcal{X}$ are the mesonic symmetries. Let us define 
$\vec{B}$ the vector of non-anomalous baryonic charges 
and $q$ the generic chemical potential associated both to the mesonic and the baryonic 
symmetries. It is natural to guess that the HS for $\firr{1}$ admits a 
decomposition in sectors having a different set of baryonic charges:
\begin{equation}\label{g1intro}
H(q;\firr{1})= \sum _{\vec{B}} m(\vec{B}) g_{1,\vec{B}}(q;\cX)
\end{equation}
where for every fixed values of $\vec{B}$ the function $g_{1,\vec{B}}(q;\cX)$ counts all the 
operators with fixed non-anomalous baryonic charges and all the possible mesonic charges; 
and $m(\vec{B})$ are some multiplicities: the number of different operators with the 
same values all non-anomalous charges.
If we consider only the non-anomalous baryonic charges the multiplicity $m(\vec{B})$ 
is in general a complicated function, if instead we consider also the anomalous 
charges the multiplicity is always one. The non-anomalous charges are of course the good 
quantum variables, but we will see that the addition of the anomalous charges will be useful to
 study interesting properties of the non-abelian spectrum and of the non-abelian moduli 
space.

One can verify that the HS actually decomposes as in (\ref{g1intro}). We can compute it
in at least two ways. We can calculate the left end side of (\ref{g1intro}) using the field theory
and some algebraic geometric techniques, and then perform some king of generalized Laurent 
expansion in the baryonic charges \cite{Forcella:2007wk,Butti:2007jv}. Alternatively we can compute 
it directly on the gravity side \cite{Martelli:2006yb,Benvenuti:2006qr,Butti:2006au,Butti:2007jv}. 
To every BPS operators 
one can associate a quantum state of a D3 brane wrapped 
over a three cycle in H. If the three cycle is trivial the corresponding 
operator will be a mesonic operator; 
if the three cycle is non-trivial the associated operator is a baryonic 
operator. There exist an infinite number of states of D3 branes wrapped in H. Indeed every section
 of every possible line bundle over $\cX$ individuates a 
suspersymmetric embedding of a D3 brane. The topology of the cycle fixes the baryonic charges $\vec{B}$ 
and using an index theorem we can compute $g_{1,\vec{B}}(q;\cX)$ counting 
all the sections of the associated line bundle 
according to their charges under the isometry of $\cX$. The multiplicities $m(\vec{B})$
can be computed instead using the Kahler cone and the quiver combinatoric. 
We address the interested reader 
to the original papers, here we just want to point out that (\ref{g1intro})
 can be computed for every SCFT on D3 branes at singularities both in field theory 
and in the dual gravity side, and the two results agree.

$H(q;\firr{1})$ contain very interesting informations:
\begin{itemize}
\item{$g_{1,0}(t;\cX)$ is the mesonic partition function, it contains the structure of 
the algebraic equations defining the variety $\cX$ and part of the KK spectrum of the 
supergravity reduction over H;}
\item{taking a limit and minimizing $g_{1,\vec{B}}(t; \cX)$, for generic $\vec{B}$, we can obtain the volume of H, and for $\vec{B}\ne 0$ the volumes of all the possible non-trivial
 three cycles $C_3$ in H. Moreover using the AdS/CFT correspondence 
we obtain the value of the central charge a and of the R-charge for every 
elementary field at the strong coupling fixed point.}
\end{itemize}


\section{$N>1$: Hilbert Series, Master Space and Plethystic Exponential}
  
We would like now to discuss the BPS operators in the non-abelian case. The step from 
N=1 to generic N is rather non-trivial due to the non-abelian dynamics and the non-trivial 
finite-N relations among the BPS operators. Moreover a direct computation of the HS in the 
field theory side becomes computationally prohibitive around N=3. In this section we want to 
point out that the knowledge of the geometrical structure of $\firr{1}$, namely 
its symmetries and its HS decomposition (\ref{g1intro}), is enough to obtain the HS counting
all the local BPS operators for generic N and to obtain some interesting informations
about the non-abelian moduli space $\CM_N$ \cite{Butti:2007jv,Forcella:2008bb}. 

As we have already explained the geometric intuition for the D3 brane system induce 
to think that to pass from N=1 to generic N we need to consider some sort of 
N-times symmetric product. 
The direct analysis of the generic BPS operator and the geometric quantization of the dual 
D3 brane state in presence of N units of flux, support this guess 
and define the precise meaning of the N-times symmetric product.

The moduli space $\CM_N$ is non-toric and non-CY. It can be described as
a $\mathbb{C}^*$ fibration over the mesonic moduli space: 
$\CM_N \simeq (\mathbb{C}^*)^{g-1} \rightarrow Sym^N \cX$. 
To explicitly write down the generic form of the HS of $\CM_N$, 
it is useful to introduce a special function called
the Plethystic Exponential: PE$[...]$. It is the generating function 
for the order N-symmetric products. Indeed
the Plethystic Exponential (PE) evaluated on a function 
$g_1$, counting a certain set of operators, generates 
the functions $g_N$ counting all the possible N-times symmetric 
product of the operators counted by $g_1$.   

Let us introduce the chemical potential $\nu$, we define the plethystic exponential PE$_{\nu}$ as: 
\begin{equation}
\hbox{PE}_{\nu}[ g_1(t)] = \exp\left( \sum\limits_{r=1}^\infty\frac{\nu^r g_1(t^r)}{r}\right) = \sum\limits_{N=0}^\infty g_N(t) \nu^N.
\end{equation}

Using the PE we will easily pass from N=1 D brane to generic N D branes. Moreover the PE has an inverse 
function called the Plethystic Logarithm (PL$=$\hbox{PE$^{-1}$}): acting with the \hbox{PE$^{-1}$} 
on a generating function we obtain the generating series for the generators and the relations in the chiral ring.

To pass to the generic N case we just need to construct the partition function counting
all the possible N times symmetric products of the functions counted by $g_{1,\vec{B}}(t; \cX)$. 
The PE plays exactly this role. 

The Hilbert series counting the BPS operators for arbitrary number N 
of branes is obtained from (\ref{g1intro}) applying the 
PE to every fixed $\vec{B}$ sector:
\begin{equation}\label{NHS}
\sum_{\nu=0}^{\infty} \nu^N H(q;\CM_N)=\sum _{\vec{B}} m(\vec{B}) PE_{\nu}[ g_{1,\vec{B}}(t;\cX) ]
\end{equation} 
where $H(q;\CM_N)$ is the Hilbert series counting the BPS operators for N D3 branes.
Once we have $H(q;\CM_N)$ we can expand it in power of the chemical potential $q$ and obtain the 
numbers of independent local BPS operators for every values of the charges. We could also 
look for large charges limits to study thermodynamical properties of the quiver theory 
\cite{Feng:2007ur}, or we can apply the PL to $H(q;\CM_N)$ and study the structure of 
generators and relations in the chiral ring \cite{Forcella:2007wk}.
 
Summarizing: the study of the MS for just one brane 
gave us the tools to compute 
partition functions counting BPS operators for 
arbitrary number N of branes for 
every SCFT living on N D3 branes at the tip of $\cX$.

The nice surprises are not finished yet. The BPS operators 
are the holomorphic functions over the 
moduli space $\CM_N$ and knowing the complete spectrum of BPS 
operators is equivalent to know the full
moduli space. Following this line of thinking it is clear that
 $H(q;\CM_N)$ can teach us 
something about $\CM_N$. Once again the MS can help to 
obtain informations 
about $\CM_N$.

Sometimes the MS $\firr{1}$ for just one D3 brane has some symmetries 
not manifest in the UV Lagrangian describing the field theory; we call them hidden symmetries. 
These symmetries are typically non-abelian extensions 
of some anomalous or non-anomalous baryonic $U(1)$ symmetries. 

It is nice to observe that the symmetries of the moduli space $\CM_N$ can be naturally 
studied with the help of the HS (\ref{NHS}). 
Indeed sometime it is possible to reorganize the spectrum of BPS operators 
in multiplets of the hidden symmetries in such a way that the HS for the $N=1$ 
master space will be explicitly written in term of characters $\chi_{\vec{l}}~(p)$ 
of these symmetries. 
If it is the case, it would imply that the BPS operators 
are in representation of the symmetries of $\firr{1}$. 
We just learned how to pass from one brane to generic N. If we are able to reorganize the 
generating functions $H(q;\CM_N)$, for the non-abelian case, 
in representations of the global symmetries guessed form the abelian case, we could claim 
that both the BPS operators and the complete moduli space $\CM_N$, for generic $N$, have the 
hidden symmetries of the N=1 Master Space $\firr{1}$. 

Indeed in a case by case analysis it is possible to show that, if 
we introduce a set of chemical potentials $p_i$, 
$i=1,...,g+2$ parameterizing the Cartan sub-algebra of the hidden 
symmetries group, it is sometime possible to rewrite the HS of $\CM_N$ 
in such a way that it will have a nice expansion in characters:
\begin{equation}
H(p;\CM_N)=\sum_{\vec{l}}\chi_{\vec{l}}~(p)
\end{equation}

We could try to use the tools we just developed to study other problems, and we will comment 
about some applications in the conclusions. Right now we would like to give an easy explicit example 
to illustrate some of the general points explained in the main text.

\section{An Example: the Cone over the Zeroth Hirzebruch Surface}

\begin{figure}[h]
\centering
\includegraphics[scale=0.7]{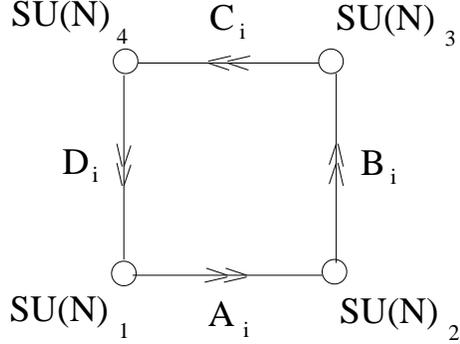}
\caption{The quiver for the $\mathbb{F}_0$ field theory.}
\label{F0}
\end{figure}

Till now the discussion was quite abstract, let us try to give an explicit example.
We choose to study the low energy field theory living on N D3 branes at the tip of the complex cone over 
$\IP^1 \times \IP^1$. 
This particular surface is called the zeroth Hirzebruch surface $\mathbb{F}_0$. 
The field theory can be nicely represented by the quiver diagram in Figure \ref{F0}. To every 
small circle we associate an SU(N) gauge group factor, and to every arrow from SU(N)$_i$
to SU(N)$_j$ we associate a chiral field $X_{ij}$ charged under the foundamental of SU(N)$_i$ ad the anti-foundamental of SU(N)$_j$.
The superpotential of the theory is:
$W = \epsilon_{ij}\epsilon_{pq} {A}_i{B}_p{C}_j{D}_q$. 
Let us start with the abelian N=1 case. 
The MS $\f_1$ can be readily found taking the derivatives of the superpotential.
There are three irreducible pieces:
\begin{equation}\label{F0I}
\f_{1} = \firr{1} \cup L^1 \cup L^2, \\
\end{equation}
with
\begin{equation}\label{FF0}
\begin{array}{rcl}
\firr{1}  &=& \mathbb{V}(B_2 D_1 - B_1 D_2, A_2 C_1 - A_1 C_2)\\
L^1 &=& \mathbb{V}(C_2, C_1, A_2, A_1)\\
L^2 &=& \mathbb{V}(D_2,D_1, B_2, B_1) \ .
\end{array}
\end{equation}
where $\mathbb{V}(...)$ means the locus of zero of the polynomials inside the brackets.
The MS has a rich structure. It is reducible
in three irreducible components. The $\firr{1}$ is $g+2=4+2=6$ dimensional 
and it is a cone. The complete intersection $xy=wz$ is a well known toric CY three-fold, it is called
the conifold and it has $SU(2)\times SU(2)\times U(1)$ isometry group. 
$\firr{1}$ is the product of two conifold and it is indeed a six dimensional CY toric 
cone. 
If we give weight $t$ to all the eight basic fields in the quiver the HS of $\firr{1}$ is:
\begin{equation}\label{HSF0}
H(t;~\firr{1})=\frac{(1 + t)^2}{(1-t)^6} \ .
\end{equation} 
Let us analyze the symmetries of the field theory. From the form of the quiver and of 
the superpotential we can see that there are two mesonic non-abelian $SU(2)$: the fields 
$A$ and $C$ are doublets of the first $SU(2)$, while $B$ and $D$ are doublets of the second
$SU(2)$. On top of these there are also a non-anomalous $U(1)_R$ R-symmetry under which all the 
elementary fields have the same $1/2$ charge, and a non-anomalous $U(1)_B$ baryonic symmetry with
charge +1 for the fields $A$ and $C$, and -1 for the fields $B$ and $D$.
At the classical level there are other two $U(1)$s that are actually anomalous in the quantum
theory. We conclude that the classical symmetry of the field theory is:
$SU(2)^2\times U(1)^4$. The $\firr{1}$ is the product of two conifold 
and has $SU(2)^4\times U(1)^2$ symmetry.
Namely the symmetry of the MS is bigger than the symmetry of the Lagrangian describing 
the SCFT. We call the new symmetries the hidden symmetries of the theory. 
 
We would like to decompose the HS of $\firr{1}$ in sectors with different baryonic charges as in 
(\ref{g1intro}). 
We introduce the chemical potentials $t$ and $b$ for the R and non-anomalous baryonic charges, 
and we define the characters $[n]\times[m]\times[p]\times[q]=[n,m,p,q]$ for the $SU(2)^4$ 
representations: the first two brackets for the $SU(2)^2$ mesonic symmetries, 
and the second two brackets for the $SU(2)^2$ hidden symmetries.  
We can organize the four fields $A,C$ in the $[1,0,1,0]$ representation and the four fields
$B,D$ in the $[0,1,0,1]$ representation. The HS (\ref{HSF0}) can be refined with all the 
chemical potentials and can be written explicitly in $SU(2)^4\times U(1)^2$ covariant form:

\begin{equation}
H (q;~\firr{1}) =  \sum\limits_{\beta, \beta' = 0}^\infty [\beta, \beta' , \beta, \beta' ] t^{\beta + \beta'} b^{\beta - \beta'} 
\label{HSF0ref}
\end{equation}

To write down the decomposition (\ref{g1intro}) we just need to explicitly write the 
$SU(2)^4$ characters and set to one the chemical potentials associated to 
the $SU(2)^2$ hidden symmetry group. This last operation would generate the multiplicities
$m(\vec{B})$.

Using the completely refined decomposition of the HS for $\firr{1}$ we can obtain the expression
for the HS counting all the local BPS operators for generic N, according to their symmetries.
The final result is:
\begin{eqnarray}
\sum_{\nu=0}^{\infty} \nu^N H(q;\CM_N) &= &
\sum\limits_{\beta,\beta' = 0}^\infty [0, 0, \beta, \beta'] \left ( PE_{\nu} \left [ \sum\limits_{n=0}^\infty [2 n + \beta, 2 n + \beta', 0, 0 ] t^{4 n + \beta + \beta'} b^{\beta - \beta'} \right ] \right. \nonumber \\
&& \left.
 - PE_{\nu} \left [ \sum\limits_{n=1}^\infty [2 n + \beta, 2n +\beta', 0, 0] t^{4 n +  \beta + \beta'} b^{\beta -\beta'} \right ] \right ) .
\end{eqnarray}
Note that the first PE contains all the terms in the second PE and hence 
all the coefficients in the expansion are positive. This is the explicit demonstration 
that for generic N the chiral spectrum organizes into representations of 
$SU(2)^4 \times U(1)^2$ and, as a consequence, also the moduli space of the non-abelian theory 
with generic rank N has symmetry $SU(2)^4 \times U(1)^2$. 
Once again if we want to count the operators according to the non-anomalous symmetries as
in (\ref{NHS}) it is enough to explicitly write down the characters and then set 
to one the chemical potentials associated to the hidden symmetries.

\section{Conclusions}

We tried an alternative algebraic geometrical approach to four dimensional 
$\mathcal{N}=1$ supersymmetric field theories. We make 
strong use of the concept of Master Space, Hilbert Series and Plethystic Exponential. 
For the specific setup of D3 branes at CY singularities we got as a result the 
control over the spectrum of BPS operators and its symmetry properties for generic N; 
the complete description for the moduli space for one brane, and interesting properties for generic N.

There are many other applications or generalizations of the techniques and results we discussed in 
these notes. For example it is possible to extend the counting procedure to include 
also the fermionic operators \cite{Kinney:2005ej,Forcella:2007ps}, 
to study marginal deformations of the field theories \cite{Butti:2007aq}
or to study the behavior of HS and MS under quantum dualities like Seiberg duality \cite{Forcella:2008ng}.

It is important to observe that, even if the tools we developed work very well 
 for the gauge theories living on D3 branes 
at CY three-fold, they can easily be applied to other interesting field theories.
Some examples are the study of SQCD-like theories \cite{Gray:2008yu} or, 
more recently, the study of the three dimensional conformal field theories living on 
M2 branes at CY four-fold singularities \cite{Hanany:2008cd,Hanany:2008qc}.

\acknowledgments
I would like to thanks my collaborators for the nice feeling all 
along these years and the INFN and SIF that gave me the opportunity 
to write this short review.
D. ~F.~ is supported by CNRS and ENS Paris.

\end{document}